\documentclass[aip,jmp,reprint,amsmath,amssymb,superscriptaddress,longbibliography,eqsecnum,nofootinbib]{revtex4-2}

\usepackage{graphicx,bm,color,mathptmx,newtxtext,newtxmath}

\usepackage[colorlinks=true,linkcolor=blue,citecolor=blue,urlcolor =blue]{hyperref}
\newcommand{\Tr}{\mathop{\mathrm{Tr}} \nolimits}

\newcommand{\sugg}[1]{{#1}}

\begin{document}

\title{Local sampling of the SU(1,1) Wigner function}

\author{Nicolas~Fabre}
\affiliation{Departamento de \'Optica, Universidad Complutense, 28040~Madrid,~Spain}
\affiliation{Telecom Paris, Institut Polytechnique de Paris, 91120~Palaiseau, France}

\author{Andrei~B.~Klimov}
\affiliation{Departamento de F\'{\i}sica, Universidad de Guadalajara, 
44420~Guadalajara, Jalisco, Mexico}

\author{Gerd~Leuchs}
\affiliation{Max-Planck-Institut für die Physik des Lichts, 91058~Erlangen, Germany}
\affiliation{Institut f\"{u}r Optik, Information und Photonik, Friedrich-Alexander-Universit\"{a}t Erlangen-N\"{u}rnberg, 91058~Erlangen, Germany}

\author{Luis~L.~S\'{a}nchez-Soto}
\affiliation{Departamento de \'Optica, Universidad Complutense, 28040~Madrid,~Spain}
\affiliation{Max-Planck-Institut für die Physik des Lichts, 91058~Erlangen, Germany}

\begin{abstract}
Despite the indisputable merits of the Wigner phase-space formulation, it has not been widely explored for systems with SU(1,1) symmetry, as a simple operational definition of the Wigner function has proved elusive in this case. We capitalize on the unique properties of the parity operator, to derive in a consistent way a \emph{bona fide} SU(1,1) Wigner function that faithfully parallels the structure of its continuous-variable counterpart. We propose an optical scheme, involving a squeezer and photon-number-resolving detectors, that allows for direct point-by-point sampling of that Wigner function. This provides an adequate framework to represent SU(1,1) states satisfactorily. 
\end{abstract}

\maketitle

\section{Introduction}

The phase-space formulation of quantum theory~\cite{Tatarskii:1983uq,Hillery:1984oq,Balazs:1984cr,Lee:1995tg,Kim:1991tr,Schroek:1996fv,Ozorio:1998aa,Schleich:2001hc,QMPS:2005mi,Weinbub:2018aa,Rundle:2021wd} stands as a self-contained alternative to the conventional Hilbert-space formalism. In this approach, observables become $c$-number functions instead of operators and quantum mechanics appears as a statistical theory. Moreover, it is the most convenient construct for visualizing quantum states and processes.

The foundations of the method were laid by Weyl~\cite{Weyl:1927aa} and Wigner.~\cite{Wigner:1932uq} Subsequently, Groenewold~\cite{Groenewold:1946aa} and Moyal~\cite{Moyal:1949fk} developed all the indispensable tools that have evolved into an accomplished discipline with applications in many diverse fields.

The phase-space picture of systems described by continuous variables, such as Cartesian position and momentum of a harmonic oscillator, gained popularity among the quantum optics community, mostly influenced by the authoritative work of Glauber.~\cite{Glauber:1963aa} In particular, the celebrated quasiprobability distributions,  such as the $P$ (Glauber-Sudarshan),~\cite{Cahill:1969aa,Sudarshan:1963aa} $W$ (Wigner),~\cite{Wigner:1932uq} and $Q$ (Husimi)~\cite{Husimi:1940aa} representations, are nothing but the functions connected with the density operator. 

The formalism has been extended in a natural way to other dynamical symmetries (recall that a Lie group $G$, with Lie algebra $\mathfrak{g}$, is a dynamical symmetry if the Hamiltonian of the system under consideration can be expressed in terms of the generators of $G$; that is, the elements of $\mathfrak{g}$).  Perhaps, the most significant example is that of SU(2), with the Bloch sphere as the underlying phase space,~\cite{Stratonovich:1956aa,Berezin:1975aa,Varilly:1989bh} which is of significance in dealing with two-level systems.~\cite{Agarwal:1981zr, Dowling:1994sw,Nieto:1998cr,Heiss:2000aa,Chumakov:2000aa,Klimov:2017aa} Entrhalling results have also been reported for the Euclidean group E(2), now with the cylinder as phase space:~\cite{Mukunda:1979uq,Plebanski:2000fk,Rigas:2011by,Kastrup:2016aa} this is of relevance in treating, e.g., the orbital angular momentum of twisted photons.~\cite{Molina:2007kn,Franke-Arnold:2008sw}  Additional developments for more general symmetries have appeared in the literature.~\cite{Brif:1998qf,Mukunda:2005aa,Klimov:2010aa,Tilma:2016aa} Moreover, the basic ideas have been adapted to discrete qudits, where the phase space becomes a finite grid.~\cite{Wootters:1987kl,Galetti:1988ff,Galetti:1992pi,Gibbons:2004ye,Vourdas:2007tw,Bjork:2008ab,Klimov:2009aa}

It is apparent from the previous discussion that a satisfactory description of a physical phenomenon requires the use of a suitable geometric arena. Surprisingly, the phase-space description of systems with SU(1,1) dynamical symmetry has received comparatively little attention,~\cite{Orowski:1990aa,Alonso:2002aa} even if SU(1,1) plays a crucial role in connection with what has been broadly termed as two-photon effects.~\cite{Wodkiewicz:1985aa,Gerry:1985aa,Gerry:1991aa,Gerry:1995kq} In the last years, the topic is becoming increasingly relevant because of the current interest in nonlinear interferometry~\cite{Yurke:1986yg} that promises stunning sub-shot noise capabilities.~\cite{Lawrie:2019td,Frascella:2019wh,Caves:2020tn,Ferreri:2021wx,Bello:2021vv} Experimental realizations of an SU(1,1) interferometer have been reported in optical systems~\cite{Jing:2011aa,Hudelist:2014aa}, spinor Bose–Einstein condensates,~\cite{Linnemann:2016aa,Gross:2010aa,Gabbrielli:2015aa} hybrid atom-light interferometers,~\cite{Chen:2015aa,Chen:2016aa} and in circuit quantum electrodynamics experiments.~\cite{Barzanjeh:2014aa}

Recently, a \emph{bona fide}  SU(1,1) Wigner function has been worked out.~\cite{Seyfarth:2020uj,Klimov:2021vi} It is defined in terms of two-mode fields and lives  in the two-sheeted hyperboloid as phase space.~\cite{Novaes:2004aa} This Wigner function appears as the  average value of the displaced parity operator, exactly as it happens for a single-mode field.~\cite{Royer:1977aa}  As this property has been employed for the direct sampling of the Wigner function for a quantum field,~\cite{Banaszek:1996un,Banaszek:1999aa,Bertet:2002aa,Sridhar:2014uu,Harder:2016aa} this opens the way for the  determination of the corresponding Wigner function for SU(1,1). Our technique avoids the detour via complex numerical reconstruction algorithms and  gives full experimental significance to the Wigner function, which can be used as a fundamental tool for interpreting the ultimate limits in the contemporary SU(1,1) setups. This is precisely the goal of this paper. 

For this purpose, we present an elemental scheme that performs a local sampling of the SU(1,1) Wigner function. It only requires a squeezer and photon-number-resolving (PNR) detectors. This matches with the prevoius schemes used to measure the single-mode Wigner distribution with only one PNR detector. Whereas in the latter case the displacement in phase space is performed with an almost-unity transmission beam splitter and a strong pump beam,~\cite{Paris:1996ve} the corresponding displacement in SU(1,1) is generated by the two-mode squeezer. After measuring in the two-mode Fock basis with two PNR detectors, straight postprocessing  allows the point-by-point determination of the SU(1,1) Wigner distribution. The state of the art in the field  makes this proposal feasible. Besides, there is no need  to use computationally costly reconstructions in this protocol.  

The paper is organized as follows. In Sec.~\ref{remindersection}, we review the basics of the SU(1,1) phase-space formalism needed for our goals. In Sec.~\ref{measurementsection}, we discuss a simple scheme for the local sampling of the SU(1,1) Wigner distribution and illustrate the method with several relevant examples. The imperfections and nonideal efficiency of the detectors is addressed in Sec.~\ref{sec:losses}, whereas our conclusions are summarized in Sec.~\ref{sec:conc}. 

%%%%%%%%%%%%%%%%%%%%%%%%%%%%%%%%%%%%%%%%%%
\section{The SU(1,1) Wigner function}
\label{remindersection}

To keep the discussion as self-contained as possible, we first briefly summarize the essential ingredients we shall be using,  according to the ideas developed in Ref.~\onlinecite{Seyfarth:2020uj}. \sugg{We deal with the superposition of two kinematically independent modes, whose  complex amplitudes (represented by operators $\hat{a}$ and $\hat{b}$, respectively) commute  ($[\hat{a}, \hat{b} ] = 0$).} 

In terms of these two modes, one can define the standard two-mode realization of the $\mathfrak{su}(1,1)$ algebra  
\begin{equation}
\hat{K}_{+} = \hat{a}^{\dagger}\hat{b}^{\dagger} \, , \qquad 
\hat{K}_{-}=\hat{a}\hat{b} \, , \qquad 
\hat{K}_{0}=\tfrac{1}{2}(\hat{a}^{\dagger}\hat{a}+
\hat{b}^{\dagger}\hat{b}+ \openone) \, ,
\end{equation}
with commutation relations (in units $h = 1$ throughout)
\begin{equation}
[\hat{K}_{0}, \hat{K}_{\pm}]= \pm \hat{K}_{\pm} \, , \
\qquad 
[\hat{K}_{-},\hat{K}_{+}]= 2\hat{K}_{0} \, .
\end{equation}  

Without going into unessential technical details, we recall~\cite{Bargmann:1947fk} that the irreducible representations (irreps) of SU(1,1) are labeled by the eigenvalues of the Casimir operator
\begin{equation}
  \hat{K}^{2} = \hat{K}_{0}^{2} - \hat{K}_{1}^{2} -
  \hat{K}_{2}^{2}  = k (k-1) \openone \, ,
\end{equation}
where $\hat{K}_{\pm} =  ( \hat{K}_{1} \pm i \hat{K}_{2} )$. The  irrep $k$ is carried by a Hilbert space spanned by the common eigenstates of $\hat{K}^{2}$ and $\hat{K}_{0}$: $\{ |k, \mu \rangle\,  :  \mu = k, k+1, \ldots \}$. The eigenvalue $k$ is known as the Bargmann index and determines the different series of irreps. Here, we focus on the so-called positive discrete series, for which $2k = 1, 2, \ldots$, as this the common case in quantum applications. The action of the generators $\{ \hat{K}_{0} , \hat{K}_{\pm}\}$ therein is
\begin{eqnarray}
  \hat{K}_{0}  |k, \mu \rangle &= & \mu |k, \mu \rangle \nonumber \, , \\
\\
  \hat{K}_{\pm}  |k, \mu \rangle & = & \sqrt{(\mu \pm k) (\mu \mp k \pm 1)}\,
  |k, \mu \pm 1 \rangle \, . \nonumber 
\end{eqnarray}
For fixed $k$, \sugg{the states $\{ |k, \mu \rangle \}$ form} a basis for the positive discrete series representation $k$, denoted  by~$\mathcal{D}_{k}^{+}$.

If the number of excitations in modes $a$ and $b$ are $n_{a}$ and
$n_{b}$, respectively, then $k$ and $\mu$ are given by 
\begin{equation}
\label{relationlabel}
  k= \tfrac{1}{2} (|n_{a} - n_{b}| + 1) \, ,
  \qquad
  \mu = \tfrac{1}{2} (n_{a} + n_{b} +1) \, .
\end{equation}
Conversely, any two-mode Fock state $| n_{a}, n_{b} \rangle$ can be mapped as states in an irrep of SU(1,1), although that is more involved as the state can appear in several irreps.~\cite{Joanis:2010vf}

We always can consider $n_{a} > n_{b}$, since the opposite can be obtained by just a relabelling of modes, with no physical consequences.  The total Hilbert space of the two oscillators decomposes then  as
\begin{equation}
  \label{eq:split}
\mathcal{H}_{a} \otimes \mathcal{H}_{b} =
\mathcal{D}_{\frac{1}{2}}^{+}
\oplus
\mathcal{D}_{1}^{+}
\oplus
\mathcal{D}_{\frac{3}{2}}^{+} \oplus \cdots . 
\end{equation}
Using this basis, every pure state can be decomposed in a full SU(1,1)-invariant way; viz, 
\begin{equation} 
\sugg{|\psi \rangle = \sum_{k} \sum_{\mu} \psi_{k\mu} \, |k, \mu \rangle \, ,}
\end{equation} 
with coefficients $\psi_{k\mu} = \langle k, \mu | \psi \rangle$.

Displacements in the SU(1,1) phase space are generated by the two-mode squeeze operator
\begin{equation}
\hat{S}(\zeta)=\exp (\zeta \hat{K}_{+} - \zeta^{\ast} \hat{K}_{-}) \, ,
\end{equation}
which acts on the modes as a Bogoliubov transformation:
\begin{align}
\label{newmode}
\hat{a}^{\prime} & = \hat{S}(\zeta) \; \hat{a} \; \hat{S}^{\dagger}(\zeta) =  \hat{a} \cosh (\tau/2) - \hat{b}^{\dagger}  e^{i \chi} \sinh (\tau/2) \, , \nonumber \\
& \\
\hat{b}^{\prime} & = \hat{S}(\zeta) \; \hat{b} \; \hat{S}^{\dagger}(\zeta) = \hat{b} \cosh (\tau/2) - \hat{a}^{\dagger} e^{i\chi} \sinh (\tau/2)\, , \nonumber
\end{align} 
where we have written
\begin{equation}
\zeta = \frac{1}{2} \tau e^{ i \chi} \, .
\end{equation}
With this parametrization  $\chi$ and $\tau$ can be interpreted as azimuthal and polar angles on a two-sheeted hyperboloid $\mathbb{H}_{2}$.~\cite{Hasebe:2019aa} Actually, there is a one-to-one correspondence between points in the complex plane $\zeta \in \mathbb{C}$ and the upper sheet of the hyperboloid $\mathbb{H}_{2}$, established via stereographic projection from the south pole:
\begin{equation}
\label{eq:defxi}
  \xi = \tanh (\tau/2) e^{ i \chi} \Leftrightarrow
  \mathbf{n} = (\cosh \tau, \sinh \tau \cos \chi, \sinh \tau \sin
  \chi)\, ,
  \end{equation}
  where $\mathbf{n}$ is a unit vector on $\mathbb{H}_{2}$, with the
  usual metric $\mathbf{n}^{2}= n_{0}^{2}- n_{1}^{2} -n_{2}^{2}$

In each irrep labeled by the Bargmann index $k$, the SU(1,1) parity operator is~\cite{Hach:2018aa}
\begin{equation}
\label{parityop}
\hat{\Pi}=(-1)^{\hat{K}_{0}-k} = \exp \left [ i \frac{\pi}{2} (\hat{a}^{\dagger}\hat{a}+\hat{b}^{\dagger}\hat{b} - k \openone) \right ] \, .
\end{equation}
Note that the SU(1,1) parity is not, in general, the parity of the photon numbers.  Nonetheless, it is a basic ingredient in several schemes proposed to beat the Heisenberg limit in SU(1,1) interferometry.~\cite{Anisimov:2010aa,Plick:2010aa,Gerry:2011aa,Li:2016aa}

According to the ideas in Ref.~\onlinecite{Seyfarth:2020uj}, the SU(1,1) Wigner function can be defined as the average value of the displaced parity operator, viz
\begin{equation}
\label{eq:defW}
W_{\varrho} (\zeta) = \Tr [ \hat{\varrho} \, \hat{S} (\zeta) \, \hat{\Pi} \, \hat{S}^{\dagger} (\zeta) ] \, ,
\end{equation}  
which, taking into account the properties of the squeeze operator, can be alternatively recast as $W_{\varrho} (\zeta )=\Tr [ \hat{\varrho} \hat{S}(2\zeta) \, \hat{\Pi}]$.

For a pure state, using the expansion in the $\{ |k, \mu \rangle \}$ basis, we get the simpler form
\begin{equation}
W_{\psi }(\zeta )=\sum_{k}\sum_{\mu ,\mu^{\prime}} (-1)^{\mu - k} \psi_{k\mu}^{\ast} \psi_{k\mu^{\prime}}^{\phantom{\ast}} \, d_{\mu \mu ^{\prime}}^{k}(2\tau )e^{i(\mu -\mu ^{\prime }) \chi },
\end{equation}
where the $d^{k}_{\mu^{\prime}\mu}(\tau)$ are the hyperbolic counterparts of the standard Wigner $d$ functions for SU(2);~\cite{Varshalovich:1988ct} that is,
\begin{equation}
  d_{\mu^{\prime} \mu}^{(k)} ( \tau ) = 
  \langle k, \mu^{\prime} |e^{i \tau  \hat{K}_{y}} |k, \mu \rangle \, .
\end{equation}
They can be expressed in a closed form in terms of the hypergeometric functions,~\cite{Vilenkin:1991aa,Ui:1970nh} although the explicit expression is of little use for our purposes here.   At the origin, this Wigner function reduces to  
\begin{equation}
\label{Wignerorigin}
W_{\psi}(0)= \sum_{k, \mu} (-1)^{\mu-k} \; |\psi_{k\mu}|^{2} \, .
\end{equation}
Note that, in principle, the map $\hat{\varrho} \mapsto W_{\varrho}$ should be invertible; i.e.,  $\hat{\varrho}$ can be obtained from $W_{\varrho}$. By using the orthogonality relation of the $d$-functions~\cite{Inomata:1992aa}
\begin{equation}
\int d\tau \sinh \tau \; d^{k}_{\mu\mu+\nu}(\tau) \, d^{k}_{\xi\xi^\prime} (\tau) = \frac{1}{2k-1}  \delta_{\mu\xi}\delta_{\mu+\nu \xi^{\prime}}\delta_{kk^{\prime}} \, , 
\end{equation}
we find that
\begin{equation}
\psi _{k\mu}^{\ast} \psi_{k\mu^{\prime}}^{\phantom{\ast}} = (-1)^{\mu -k} \frac{ 2k-1}{2 \pi} \int d\chi d\tau \sinh \tau \; W_{\psi}(\zeta /2) e^{i( \mu^{\prime} - \mu ) \chi} \, d_{\mu^{\prime}\mu }^{k}(\tau ) \, .
\end{equation}
Therefore, by measuring the SU(1,1) Wigner function we obtain
\begin{equation}
\hat{\varrho}_{\mathrm{WF}} = \sum_{k}\sum_{\mu ,\mu^{\prime}} \psi _{k\mu}^{\ast} \psi_{k\mu^{\prime}}^{\phantom{\ast}} \, | \mu^{\prime },k\rangle \langle \mu,k| \, .
\end{equation}
In other words, the mapping is invertible only for a single irrep, where the concept of phase space is uniquely defined. Nonetheless, since the Hilbert space splits as in \eqref{eq:split}, our Wigner function appears as a sum over all the irreps in the discrete positive series. In any case, \eqref{eq:defW} is a physically meaningful definition that allows us to visualize every quantum state of interest.  More generally, the reconstructed Wigner function allows computing only average values of any function of the SU(1,1) generators.
 
Alternatively, we can use the two-mode Fock basis, so that for a pure state $| \psi \rangle =\sum_{n_{1},n_{2}} \psi_{n_{1}n_{2}} |n_{1},n_{2}\rangle $, the SU(1,1) Wigner function can be written as
\begin{align}
\label{Wignerfock}
W_{\psi}(\zeta) = \sum_{n_{1},n_{2}} \sum_{n^{\prime}_{1},n^{\prime}_{2}} \sum_{n_{a},n_{b}} (-1)^{\case{1}{2}(n_{a} + n_{b}+1)-k}  \;
\psi^{\ast}_{n^{\prime}_{1} n^{\prime}_{2}} \psi_{n_{1}^{\phantom{\prime}}n_{2}^{\phantom{\prime}}}^{\phantom{\ast}}  \; S^{n_{a}n_{b}}_{n_{1}^{\prime}n_{2}^{\prime}}(\zeta) S^{n_{1}n_{2}}_{n_{a}n_{b}}(- \zeta)  \, ,
\end{align}
where $S^{n_{a}n_{b}}_{n_{1}^{\prime}n_{2}^{\prime}}(\zeta)=\langle n^{\prime}_{1},n^{\prime}_{2}| \hat{S}(\zeta) | n_{a},n_{b}\rangle$.~\cite{Caves:1991tg,Selvadoray:1994uj} At the origin, we have now 
\begin{equation}
\label{eq:sWn}
W_{\psi}(0)=\sum_{n_{a},n_{b}} (-1)^{\case{1}{2} (n_{a} + n_{b}+1)-k} \; |\psi_{n_{a}n_{b}}|^{2} \, ,
\end{equation} 
which is a direct generalization of the expression used in the single-mode case.

\section{Sampling the SU(1,1) Wigner function}
\label{measurementsection}

\begin{figure}[h]
 \centering
\includegraphics[height=4.5cm]{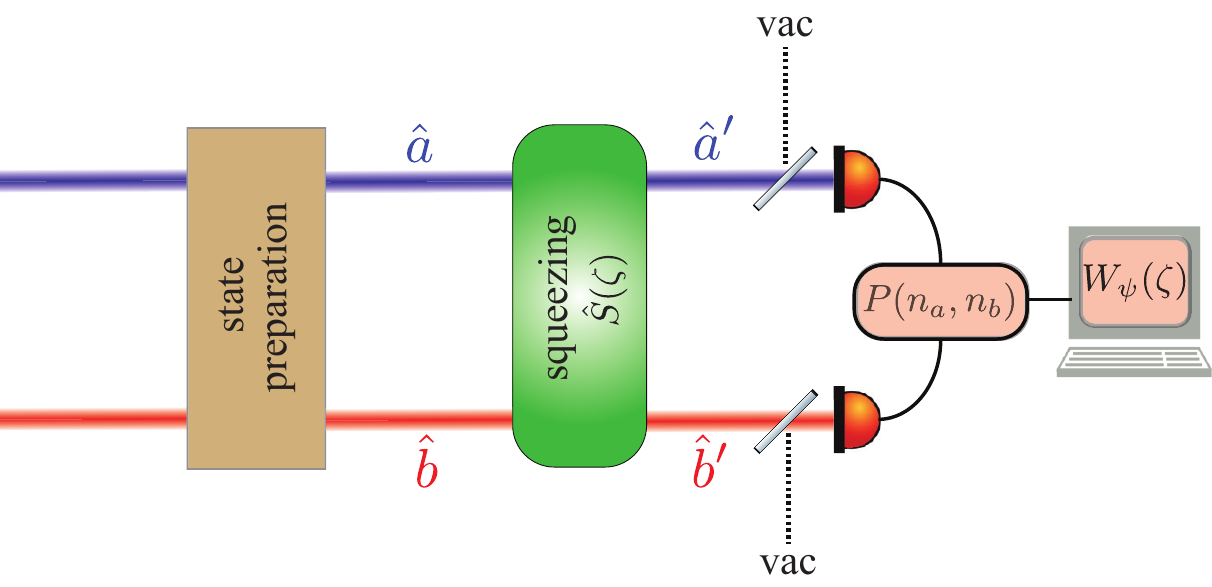}
\caption{\label{schematic} Optical scheme for sampling the SU(1,1) Wigner function. The two-mode state is first prepared and then a squeezer is applied. The joint photon-number distribution $P(n_{a}, n_{b})$ is measured with PNR detectors.  The finite efficiency of these detectors is modeled by a beam splitter.}
\label{fig:scheme}
\end{figure}

In this section, we discuss a  scheme to sample point-by-point the SU(1,1) Wigner function introduced in the previous Section. The setup is depicted in  Fig.~\ref{fig:scheme}. In a first step, we prepare a source of two-mode quantum states that, due to their symmetry properties, are most adequately described in terms of the SU(1,1) invariant basis.  
%%%%%%%%%%%%%%%%%%%%%%%
%%%%%%%%%%%%%%%%%%%%

By placing PNR detectors at each mode $a$ and $b$, we directly acquire the   histogram $P(n_{a},n_{b})=|\langle n_{a},n_{b}| \psi\rangle |^{2} = |\psi_{n_{a}n_{b}}|^{2}$, where, for simplicity, we have assumed pure states. According to \eqref{relationlabel} this is tantamount to access to  
\begin{align}
P(k,\mu) =  P \left (k= \tfrac{1}{2}(|n_{a}-n_{b}|+1), 
\mu= \tfrac{1}{2} (n_{a}+n_{b}+1) \right )  = 
\sugg{P(n_{a},n_{b})} \, .
\end{align} 
From this histogram, we directly determine  the Wigner distribution at the origin $W_{\psi}(0)$ according to either \eqref{Wignerorigin} or \eqref{eq:sWn}.

The SU(1,1) Wigner function can be obtained at other points by displacing the two-mode field $| \psi \rangle $ with a two-mode squeezer with parameter $\zeta$. This can be implemented via, e.g., four-wave mixing  in a nondegenerate parametric amplifier~\cite{Gerry:1991tg} whose Hamiltonian is 
\begin{equation}
\hat{H}=\zeta \hat{a}^{\dagger}\hat{b}^{\dagger} -\zeta^{\ast}\hat{a}\hat{b} \,.
\end{equation}
In this case, $\zeta$ depends on the nonlinear coefficient $\chi_{\mathrm{NL}}$ of the crystal and on the amplitudes of the strong pump fields $\zeta=\chi_{\mathrm{NL}} \alpha_{p1}\alpha_{p2}$.  The nonlinear coefficient being fixed, the amplitude and phase of the two pumps have to be modulated to explore the full SU(1,1) phase space. Repeating the same procedure, we determine the Wigner function at the origin for this new set of output modes:
\begin{equation}
W_{\psi^{\prime}}(0)= \sum_{n_{a},n_{b}} (-1)^{\case{1}{2}(n_{a} + n_{b}+1)-k} \, |\psi^{\prime}_{n_{a}n_{b}}|^{2} = W_{\psi}(\zeta) \, .
\end{equation}
The measured probability distribution takes the form:
\begin{equation}
|\psi^{\prime}_{n_{a}n_{b}}|^{2}= \sum_{n_{1},n_{2}} 
\sum_{n^{\prime}_{1},n^{\prime}_{2}} 
\psi^{\ast}_{n^{\prime}_{1}n^{\prime}_{2}} \psi_{n_{1}n_{2}}^{\phantom{\ast}} \; S^{n_{a}n_{b}}_{n_{1}^{\prime} n_{2}^{\prime}}(\zeta) \;S^{n_{1}n_{2}}_{n_{a}n_{b}}(- \zeta)  \, .
\end{equation}
Therefore, we can sample the Wigner function at the desired points. Please, note that it is not necessary to know the explicit form of the $S$-coefficients for the reconstruction, because they are directly measured in the setup.  

\begin{figure}[t]
 \centering
\includegraphics[height=4.5cm]{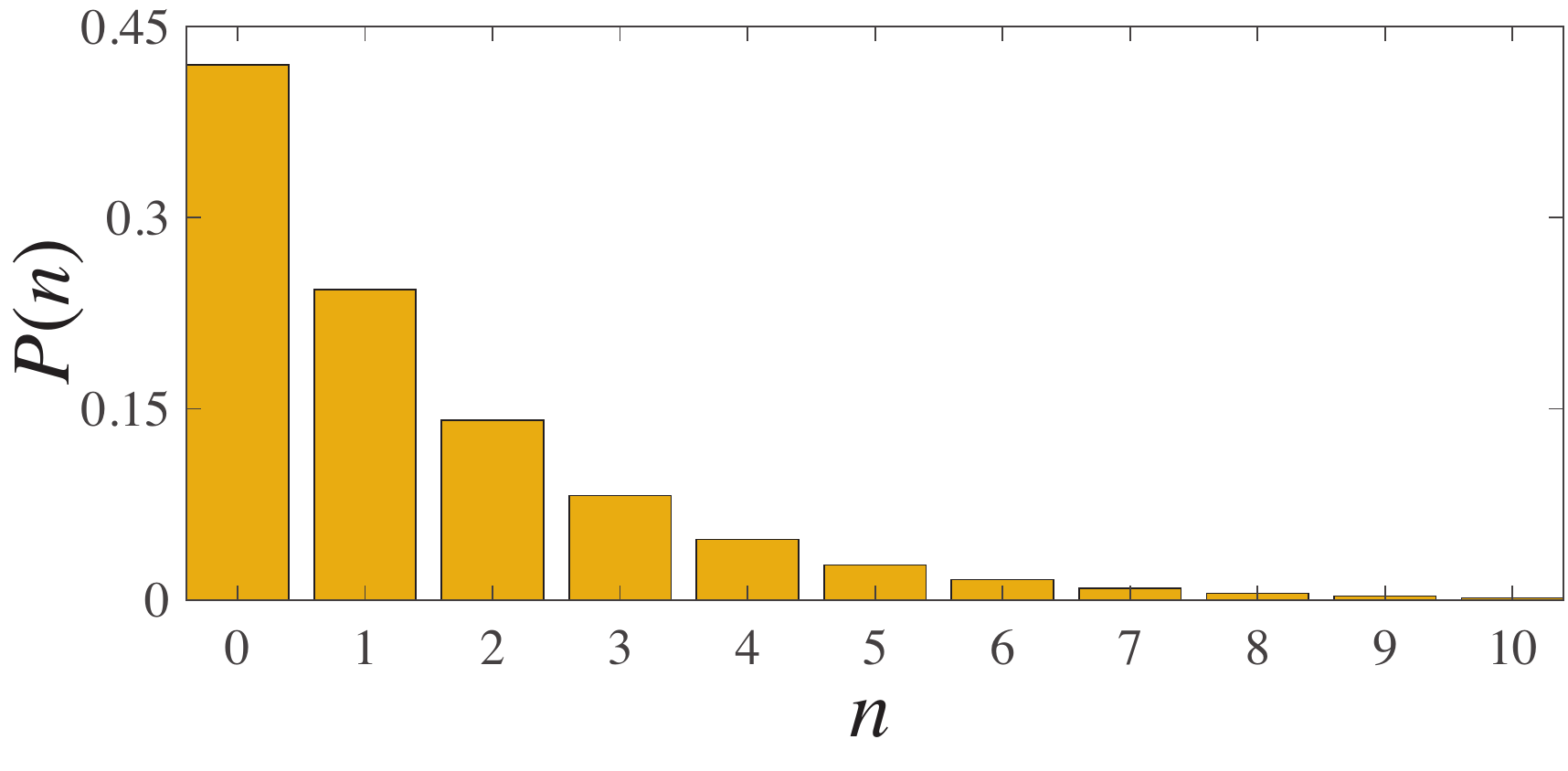}
\caption{\label{twomodehisto} Photon-number distribution $P(n)$ of a two-mode squeezed vacuum $| \mathrm{TMSV} \rangle$  with $|\zeta_{0}|= 2 \tau_{0} = 1$.}
\end{figure}

Let us illustrate our proposal with \sugg{some examples. We first take} the case of the two-mode squeezed vacuum (TMSV), defined as
\begin{align}
|\text{TMSV}\rangle & = \hat{S} (\zeta_{0} ) |0_{a},0_{b} \rangle \ = \frac{1}{\cosh (\tau_{0}/2)} \sum_{n=0}^{\infty} e^{i n \chi_{0}} \, [\tanh (\tau_{0}/2)]^{n} | n, n \rangle \, ,
\end{align}
whose  properties  have been fully explored.~\cite{Chekhova:2015vf,Oudot:2015vg} Its photon-number distribution $P(n)=|\langle n, n  |\text{TMSV}\rangle |^{2}$  (which corresponds to the probability of finding $n$ photons in each of the two modes simultaneously) is represented in Fig.~\ref{twomodehisto} with $|\zeta_{0} |= 2 \tau_{0} =1$. This distribution has been experimentally measured in Ref.~\onlinecite{Kalashnikov:2012wl}. We observe that resolving 10 photons is enough to perform the sampling. The segmented detector described in Ref.~\onlinecite{Nehra:2020wt} is able to do that, while transition-edge sensors can discriminate photon numbers up to 25 with efficiencies  higher than 87\%~\cite{Schmidt:2018wx} in the 850-950~nm range. A recent improvement  has demonstrated that, by multiplexing highly-efficient transition-edge sensors and measuring the integration time from which the material composing the detector goes from the normal to the superconducting phase after multiphoton absorption, 100 photons can be resolved.~\cite{Eaton:2022aa} With such performances, we conclude that the SU(1,1) Wigner function can be sampled.  

%%%%%%%%%%%%%%%%%%%%%%%
%%%%%%%%%%%%%%%%%%%%

\begin{figure}[b]
\centerline{\begin{tabular}{cc}
\includegraphics[width=\columnwidth]{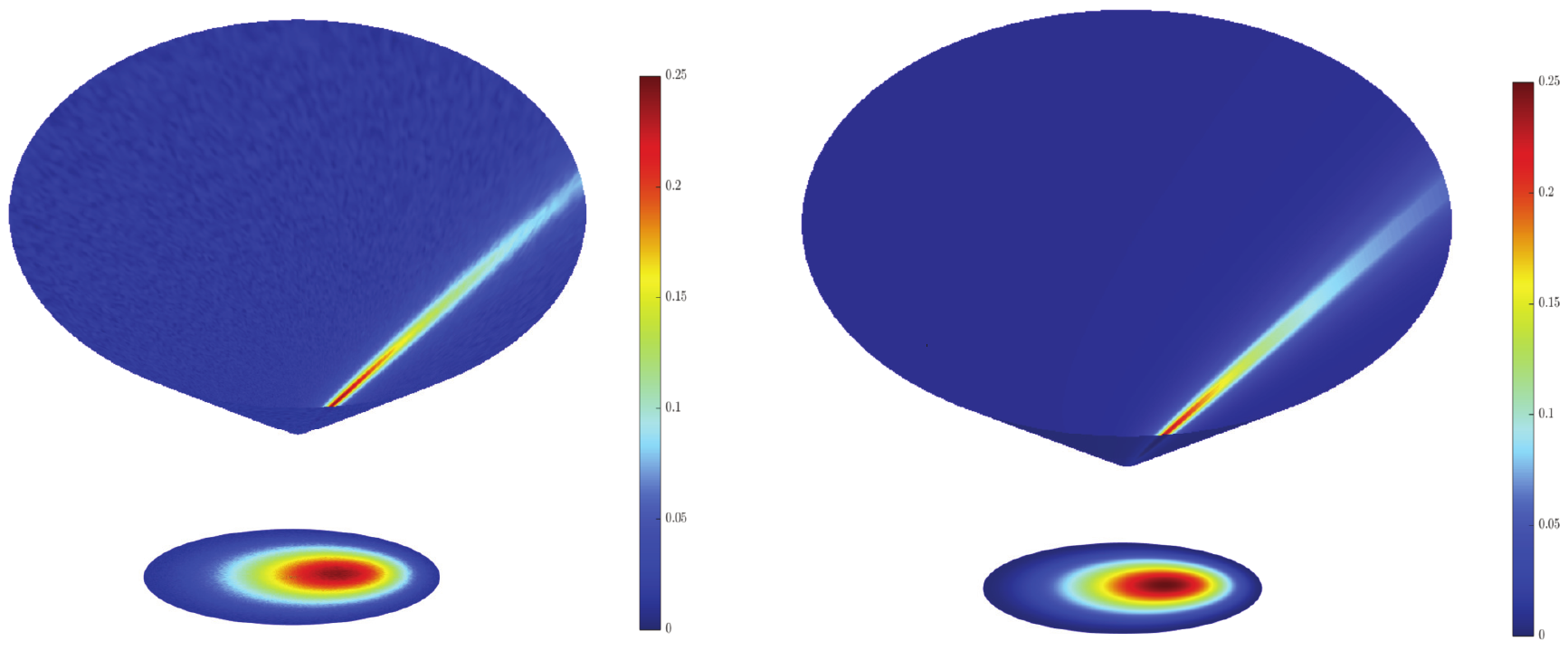} 
\end{tabular}}
\caption{\label{wignerplot} Numerical simulation of the sampled SU(1,1) Wigner function for a two-mode squeezed vacuum  with $\tau_{0}=3$ (left) and the same experiment with  Gaussian noise with a signal-to-noise of 30 (right), both as a density plot on the upper sheet of the hyperboloid, with the scales included. \sugg{In both cases we show  the associated distributions in the unit disk, obtained from the upper sheet by stereographic projection from the south pole of the hyperboloid $\mathbb{H}_{2}$.} The detectors have a resolution of 100 photons.} 
\end{figure}

Next, we consider the action of a displacement on $| \mathrm{TMSV} \rangle$; that is, the resulting state is $| \psi^{\prime} \rangle = \hat{S}(\zeta) \hat{S}(\zeta_{0})|0_{a}, 0_{b} \rangle$.  The photon-number distribution of the displaced state reads as
\begin{align}\label{twomodeproba}
P_{n}(\zeta )  = \frac{1}{4 \cosh^{2}(\tau^{\prime}/2)} \left | \frac{1+ \xi \xi_{0}^{\ast}}{1+\xi^{\ast} \xi_{0}}  \right |^{2}  
\left |   \frac{e^{i \chi_{0}} \tanh (\tau_{0}/2)
+e^{i \chi} \tanh(\tau/2)}{1+ e^{i(\chi -\chi_{0})}\tanh(\tau_{0}/2) \tanh(\tau/2)}   \right |^{2n} \, .
\end{align}
From this acquired histogram, we directly infer the Wigner function: $W_{\psi}(\zeta)=\sum_{n} (-1)^{n} P_{n}(\zeta)$. In Fig.~\ref{wignerplot}, we plot the resulting Wigner function for this two-mode squeezed vacuum in the upper sheet of the hyperboloid, the phase space for the problem.

%%%%%%%%%%%%%%%%
%%%%%%%%%%%%%%%

Let us consider the two-mode vacuum $|0_{a}, 0_{b}\rangle$, with $S$-coefficients $\langle n_{a}n_{b}|\hat{S}(-\zeta)|0_{a}0_{b}\rangle =\delta_{n_{a}n_{b}} \xi^{n_{a}}$. We thus have
\begin{equation}
\label{wignervacuum}
W_{|0_{a},0_{b}\rangle} (\zeta)=\sum_{n=0}^{\infty} (-1)^{n} |\xi|^{2n}=\frac{1}{1+|\xi|^{2}} \, ,
\end{equation}
which could be obtained from the previous case in \eqref{twomodeproba} when the squeezing parameter $\xi_{0} \rightarrow 0$.

As our last relevant example, we take the state $|1_{a}1_{b}\rangle$, the archetypical example of nonclassical light and sometimes called a biphoton.~\cite{Klyshko:1994aa} Now,  $\langle n_{a}n_{b}|\hat{S}(-\zeta)|1_{a}1_{b}\rangle =\delta_{n_{a}n_{b}}\xi^{n_{a}}[ \cosh^{-2}(\tau/2) n_{a} \xi^{-1}-\xi^{\ast}]$,  which directly leads to  
\begin{equation}
W_{|1_{a},1_{b}\rangle} (\zeta)= \frac{1}{\cosh^{2}(\tau/2)} \sum_{n=0}^{\infty} (-1)^{n} |\tanh(\tau/2)|^{2n} \left |\frac{n}{\sinh(\tau/2)}-\sinh(\tau/2)\right |^{2}.
\end{equation}

In Fig.~\ref{wignerplotother}, we have represented the Wigner functions of the two-mode vacuum and the biphoton. Both have a rotational symmetry with respect to the origin of the hyperboloid, which means that the phase of $\zeta$ is unessential to sample the distribution. While at the origin the Wigner function of the vacuum state is the unity, the one of the biphoton state is equal to minus one. \sugg{One might naively conclude that negative values indicate the presence of a non-Gaussian state. However,  the negativity of the Wigner function is merely a proxy for the non-Gaussianity of the state expressed in canonical position-momentum coordinates;~\cite{Kenfack:2004lw} this has no significance for other dynamical variables, such as the ones associated with SU(1,1), for which the very concept of Gaussianity is equivocal.~\cite{Goldberg:2020aa}}  

Given this symmetry, in Fig.~\ref{wignerplotother} we have plotted a section of both Wigner functions as a function of $\tau$.  In contrast with the two-mode squeezed state, these functions tend to one when $\tau\rightarrow \infty$ and thus spread to infinity. While the sum in \eqref{wignervacuum} involves infinite terms, the PNR detector can only resolve a finite number of photons, we denote by $N$. If we focus on the two-photon vacuum, for $|\tau|<2.3$, a PNR detector that can resolve 10 photons allows us finding the correct value of the Wigner function. For $|\tau|>2.3$, a PNR detector that can resolve around 100 photons is needed. For the biphoton, as  $N$ increases, the wings of the Wigner function are shifted.

%%%%%%%%%%%%%%%
\begin{figure}
\centerline{
\begin{tabular}{cc}
\includegraphics[width=\columnwidth]{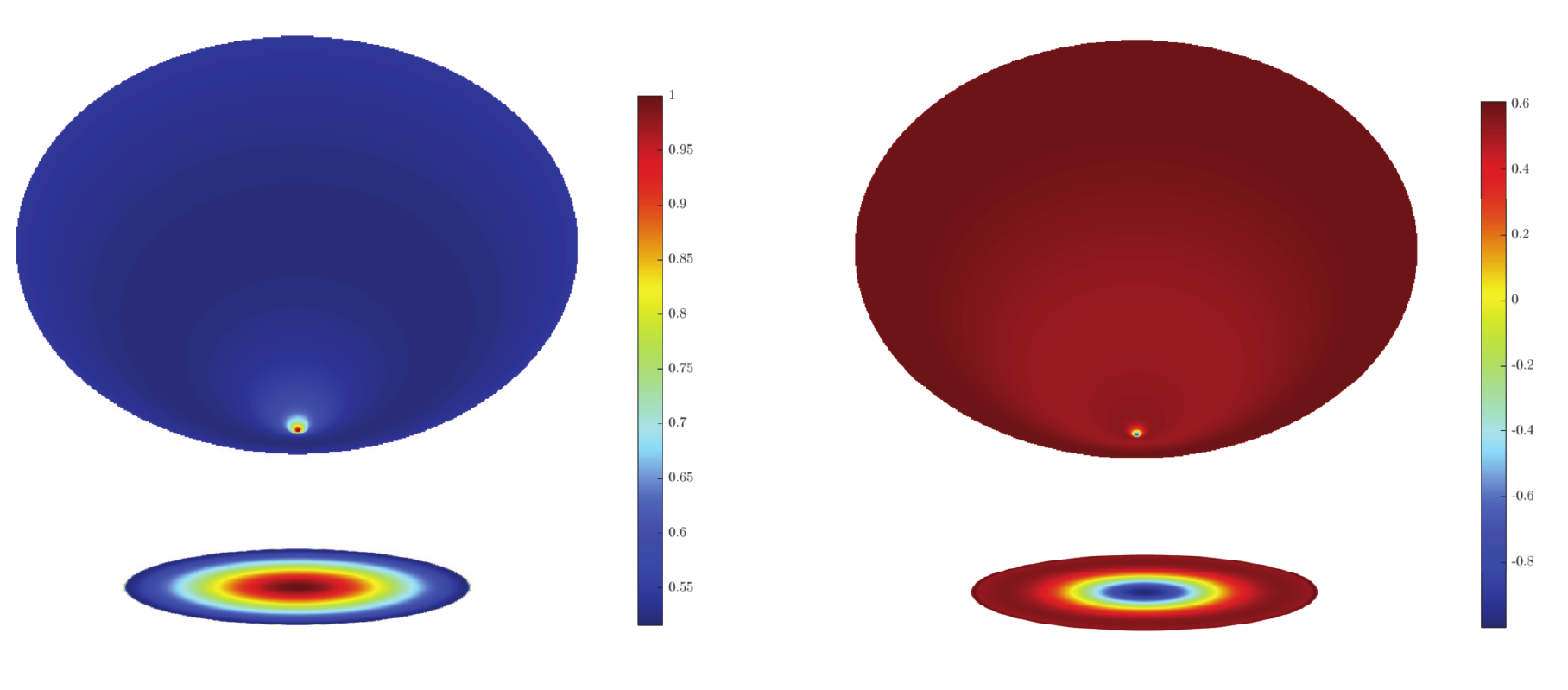}  
\end{tabular}}
\caption{\label{wignerplotother} Numerical simulation of the sampled SU(1,1) Wigner function for a two-mode vacuum state (left) and for a biphoton (right), both in the hyperboloid with the scales indicated. The Wigner function at the origin is equal to 1 (resp. $-1$) for the vacuum state (resp. the biphoton). \sugg{We also include  the associated distributions in the unit disk, obtained from the upper sheet by stereographic projection from the south pole of the hyperboloid $\mathbb{H}_{2}$.}  The detectors have a resolution of 100 photons.}
\end{figure}
%%%%%%%%%

%%%%%%%%%%%%%%%%
\begin{figure}[b]
\centerline{
\begin{tabular}{cc}
\includegraphics[width=\columnwidth]{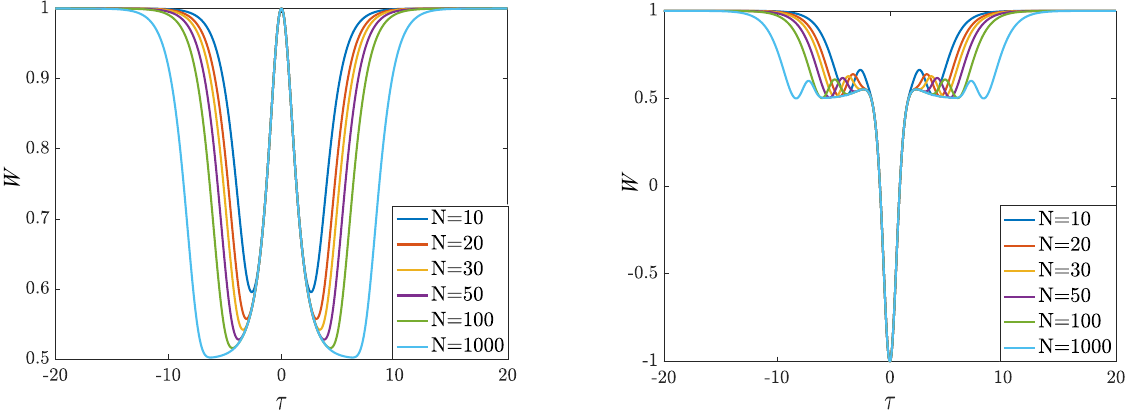} 
\end{tabular}}
\caption{\label{wignerplotother} Numerical simulation of the sampled SU(1,1) Wigner function for a two-mode vacuum state (left) and for a biphoton (right) as a function of $\tau$ for different values of the number of photons $N$ that can be resolved by the PNR detector used in sampling. }
\end{figure}

\section{Finite efficiency detectors}
\label{sec:losses}

In practice, PNR detectors always have a nonunit efficiency, meaning that even if photons are incident into the detector it has a finite probability not to trigger a click event. We denote this efficiency by $\eta$ ($0 \le \eta \le 1$). Following a standard procedure,~\cite{Rohde:2006vo} we model this situation by preceding a perfect unit-efficiency detector with a beam splitter possessing the transmittance $\eta$, as sketched in Fig.~\ref{fig:scheme}. In consequence, each mode is replaced with 
\begin{equation}
\hat{a}^{\prime} \mapsto   \sqrt{\eta}\; \hat{a}^{\prime} - \sqrt{1-\eta} \; \hat{a}_{\mathrm{vac}} \, ,
\end{equation} 
where $\hat{a}_{\mathrm{vac}}$ represents the vacuum mode. 

This transformation changes the parity operator of each individual mode to   
\begin{equation}
\hat{\Pi}_{\mathrm{vac}}= \exp \left [ -2 \eta \left ( 
\hat{a}^{\prime \dagger}-\sqrt{\tfrac{1-\eta}{\eta}}  \hat{a}^{\dagger}_{\mathrm{vac}} \right ) 
\left  ( \hat{a}^{\prime}-\sqrt{\tfrac{1-\eta}{\eta}} \hat{a}_{\mathrm{vac}} \right ) \right ] \, .
\end{equation}
By tracing over the ancillary vacuum mode, we get the single-mode parity  with finite-efficiency detectors; viz 
\begin{equation}
\label{parityafterlosses}
\hat{\Pi}_{\mathrm{loss}}= \int d\alpha :\exp \left [ -2\eta \left ( 
\hat{a}^{\dagger}-\sqrt{\tfrac{1-\eta}{\eta}} \alpha^{\ast} \right ) \left (\hat{a}-\sqrt{\frac{1-\eta}{\eta}} \alpha \right ) \right ] : \, ,
\end{equation}
where $:  \; \cdot \; :$ stands for normal ordering.

We recall the notion of $s$-ordered parity operator~\cite{Banaszek:1996un}
\begin{equation}
\label{sparityop}
\hat{\Pi}(\alpha,s)= \frac{2}{\pi(1-s)} : \exp - \left [ \frac{2}{1-s} 
( \alpha^{\ast}-\hat{a}^{\dagger})(\alpha-\hat{a}) \right ] : \,.
\end{equation}
The Wigner function at the origin is given by the expectation value of $\hat{\Pi}(0,0)$. Therefore,  
 \begin{equation}
W_{\psi}(0)=\int d\alpha  W_{\psi} \left (\sqrt{\frac{1-\eta}{\eta}} \alpha; -\frac{1-\eta}{\eta}\right ) \, , 
\end{equation}
where $(\alpha, s)$  the $s$-ordered-phase space distribution (in our case with $s=-(1-\eta)/\eta$).  

In the two-mode case, the parity operator in its normal form becomes, with   the efficiency of the two photodetectors being $\eta_{a}, \eta_{b}$:
\begin{widetext}
\begin{align}
\hat{\Pi}_{\mathrm{loss}} & = \iint d\alpha d\beta : \exp \left 
[ (i-1) \eta_{a} \left (\hat{a}^{\prime \dagger}-\sqrt{\tfrac{1-\eta_{a}}{\eta_{b}}} \alpha^{\ast} \right ) \left (\hat{a}^{\prime}-\sqrt{\tfrac{1-\eta_{a}}{\eta_{a}}} \alpha \right ) \right ]: \nonumber \\
& \times : \exp \left [ (i-1) \eta_{b} \left (\hat{b}^{\prime \dagger}-
 \sqrt{\tfrac{1-\eta_{b}}{\eta_{b}}} \beta^{\ast} \right ) \left (\hat{b}^{\prime}-\sqrt{\tfrac{1-\eta_{b}}{\eta_{b}}} \beta \right ) \right ] : \, .
\end{align}
\end{widetext}
We observe that this operator is disentangled in the modes $\hat{a}^{\prime}, \hat{b}^{\prime}$, but is entangled in the initial modes $\hat{a}, \hat{b}$. If the two modes are initially uncorrelated, $|\psi\rangle =\sum_{n_{a},n_{b}} \psi_{n_{a}} \psi_{n_{b}} |n_{a}, n_{b} \rangle$, the average value of the SU(1,1) parity operator $\hat{\Pi}_{\mathrm{loss}}$ is the product of two $s$-ordered single-mode Wigner distribution:
\begin{align}
\label{uncorrelatednoise}
W_{\psi}(0) =\int d\alpha W_{\psi_{a}} \left ( \alpha \sqrt{\tfrac{1-\eta_{a}}{\eta_{a}}}, s \right )  \int d\beta W_{\psi_{b}} \left ( \beta \sqrt{\tfrac{1-\eta_{b}}{\eta_{b}}}, s^{\prime}\right ),
\end{align}
with $s=1+\sqrt{2}e^{i\pi/4}/\eta_{a}T $ (and analogously for $s^{\prime}$).~\cite{Anaya:2019wy} The fact that $s$ is a complex value  comes mathematically from the factor two in (\ref{parityop}) or equivalently of the normal form of the two-mode parity operator. A two-mode uncorrelated case could be verified experimentally using two separable coherent states as input and once the tomography of the detectors has been performed. In the general case of correlated modes, (\ref{uncorrelatednoise}) is no longer valid. The parameter $s$ of this entangled phase-space distribution could be expressed as a function of the two efficiencies $\eta_{a}, \eta_{b}$ of the PNR detectors, but it is out of the scope of this paper. 

In Fig.~\ref{wignerplot} we plot the sampling of the Wigner distribution of the two-mode squeezed vacuum with added Gaussian noise with a signal-to-noise value of 30. Interestingly, this Wigner function is positive in the absence of noise, but when Gaussian noise is added it exhibits negative values. \sugg{This confirms that positivity with Gaussianity in the context of SU(1,1) may entail many pitfalls for the unwary.}

\section{Concluding remarks}
\label{sec:conc}

In summary, we have presented an experimental scheme to measure the SU(1,1) Wigner distribution of two-mode quantum states  that can be implemented with actual technology.  Despite the apparent difficulty of the SU(1,1) Wigner function, the presented  protocol avoids the need for a complex tomographical reconstruction algorithm.  We stress that the method is feasible with current technology and will promote this Wigner function from an academic curiosity to a realistic tool to deal with two-photon effects.

Characterization of two-mode quantum states is essential for its use as a probe in phase-estimation metrological protocols.~\cite{Adhikari:2018wd} Indeed, the SU(1,1) interferometer can beat the standard quantum limit without using quantum states as input.~\cite{Yurke:1986yg} Finally, the SU(1,1) Wigner distribution could be measured with an ancilla qubit state to implement the parity operator, based on the idea presented in Ref.~\onlinecite{Lutterbach:1997ue}.

\section*{Acknowledgments}
Our work on SU(1,1) came as a result of long conversations during the  frequent visits of Jonathan Dowling to Erlangen. Through his outstanding scientific work, his kind attitude, and his inimitable humor, he leaves behind a rich legacy for all of us. We dedicate this paper to his memory.

We thank H. de Guise, O. Pfister, U. Seyfarth and Ch. Silberhorn for discussions. This work received funding from the European Union’s Horizon 2020 research and innovation programme under grant Agreement No. 899587. We acknowledge support from the Spanish Ministerio de Ciencia e Innovación (PGC2018-099183-B-I00). A. B. Klimov acknowledges support from the Grant 254127 of CONACyT (Mexico).

\section*{Data availability}
\noindent Data sharing is not applicable to this article as no new data were created or analyzed in this study.

\section*{Conflict of interest}
\noindent The authors have no conflicts to disclose.
  
%\bibliography{angular}

%aipnum4-2.bst 2019-01-14 (MD) hand-edited version of apsrev4-1.bst

\end{document}